# From Dynamic to Lexical: A Comparative Exploration of Scoping Rules in SAS and R

Chen Ling[1] and Yachen Wang[1]

[1]AbbVie Inc., Lake County, Illinois, USA

Corresponding author: Chen Ling (ling0152@umn.edu)

**Abstract**

Variable scoping dictates how and where variables are accessible within programming languages, playing a crucial role in code efficiency and organization. This paper examines the distinct scoping rules in SAS and R, focusing on SAS's dynamic scoping and R's lexical scoping. In SAS, dynamic scoping utilizes symbol tables, resolving variables at runtime by dynamically searching through active macro layers. R, in contrast, employs lexical scoping, using environments to resolve variables based on the structure in which functions are defined.

Illustrative examples highlight the differences between these scoping strategies, showcasing their impact on code behavior. Additionally, the paper outlines methods for inspecting variables in SAS's symbol tables and R's environments, offering practical insights for debugging and optimization. Strategies for controlling variable scope in both languages are discussed, enhancing code precision and reliability. This exploration equips programmers with critical understanding to optimize variable management, improving their programming practices in SAS and R.

## 1. Introduction

Variable scoping is a key concept that influences how and where variables can be accessed within a codebase. This concept is particularly significant in functional programming, where constructs like macros in SAS and functions in R heavily rely on well-defined scoping rules to ensure accuracy and efficiency. As programming languages offer different scoping mechanisms, understanding these differences is essential for developers aiming to write precise and reliable code.

Despite its importance, there have been very few discussions regarding the differences in scoping rules between SAS and R, two widely used languages in the pharmaceutical industry that offer distinct approaches to scoping. SAS employs dynamic scoping, relying on symbol tables to manage variable resolution at runtime. This method involves searching through layers of active macro calls to find variables, which can offer flexibility in variable management. While R, on the other hand, uses lexical scoping (Wickham, 2019), with environments determining variable resolution based on the location where functions are defined. This approach offers structural stability and predictability in variable behavior.

With many organizations transitioning from SAS to R (Ling & Wang, 2025), understanding these differences in scoping rules is important for ensuring accurate programming outcomes. Through detailed examples, we explore the implications of dynamic versus lexical scoping, offering insights into their effects on code behavior. Additionally, strategies for inspecting and controlling variable scopes in both languages are discussed. This paper aims to shed light on how these differing scoping rules can impact program results, thereby facilitating a smoother transition for programmers moving from SAS to R.

## 2. What is variable scoping

Variable scoping is a fundamental concept in functional programming that determines where variables can be declared, modified, and accessed within a codebase. It defines the rules that govern the visibility and lifespan of variables, which can significantly affect a program's design and behavior. Understanding scoping is crucial for creating accurate and efficient code, as it controls how functions and macros refer to variables that are used but not defined within them.

In both SAS and R, there are generally two types of scope: Global Scope and Local Scope. As illustrated in Figure 1 below, global variables are defined in the global scope, while variables defined within macros or functions are in local scope. Global variables remain accessible throughout the entire SAS/R session, whereas local variables are confined to their specific scopes and are discarded once those scopes end. In the example below, to calculate the value of z in the local scope, we need value from both x and y. Since variable x is not defined within the local scope in both SAS and R, it is retrieved from the global scope. However, the mechanism for this "retrieval" (scoping rules) differs between the two languages, which will be discussed further in this paper.

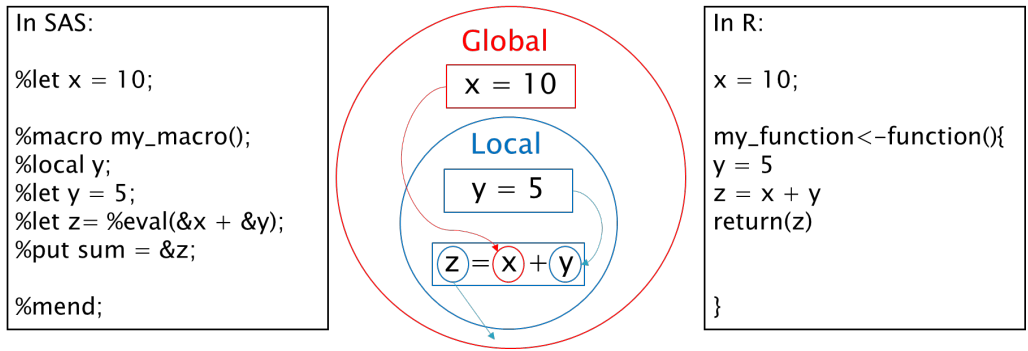

**Figure 1. Example of variable scoping in SAS and R**

## 3.     What powers variable scoping

In SAS, variable scoping relies on a special memory area called the symbol table (Buchecker, 2020), which manages macro variables and is crucial for dynamic code substitution. The symbol table contains two main types: the global symbol table and the local symbol table, corresponding to the global and local scopes mentioned earlier. There is only one global symbol table created when the SAS session starts and is deleted when the SAS session ends. However, various local symbol tables are temporarily created during macro execution under specific conditions, then they will be deleted after the execution.

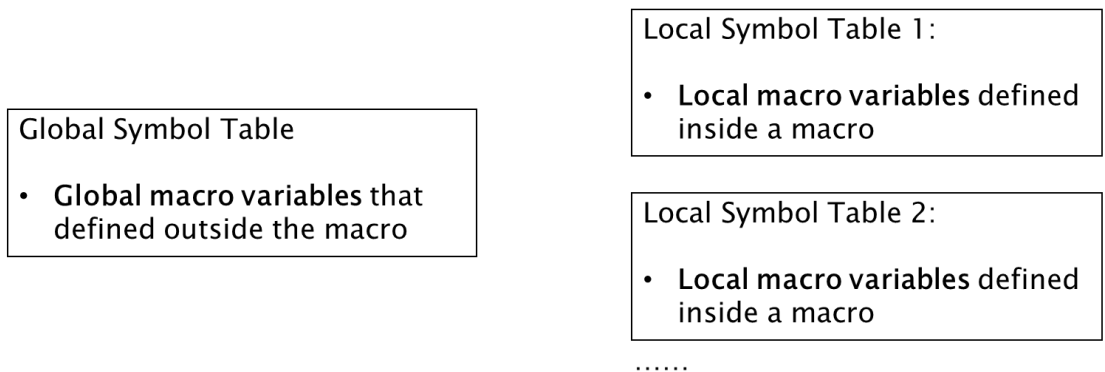

**Figure 2. Demonstration of symbol tables in SAS**

Similar to SAS symbol tables, R uses a memory area called an environment to manage variables and their values (Ling & Wang, 2025). Instead of the global and local symbol tables, in R we have global and local environments for global and local scopes. In Figure 3 below, you'll see the source code on the left and a demonstration of the code execution on the right.

- In the first step, R creates two objects in the global environment: it assigns the value 6 to y, and a function to h (depicted with a little pink symbol).

- In the second step, function h() is invoked with h(1), which initiates an execution environment for h(). Within this environment, the variable a is set to 2, and x defaults to 1.

- In the third step, the function computes the sum of x and a, returning this result to be assigned to z in the global environment. Once the function h() completes its execution and returns the value, the execution environment is discarded.

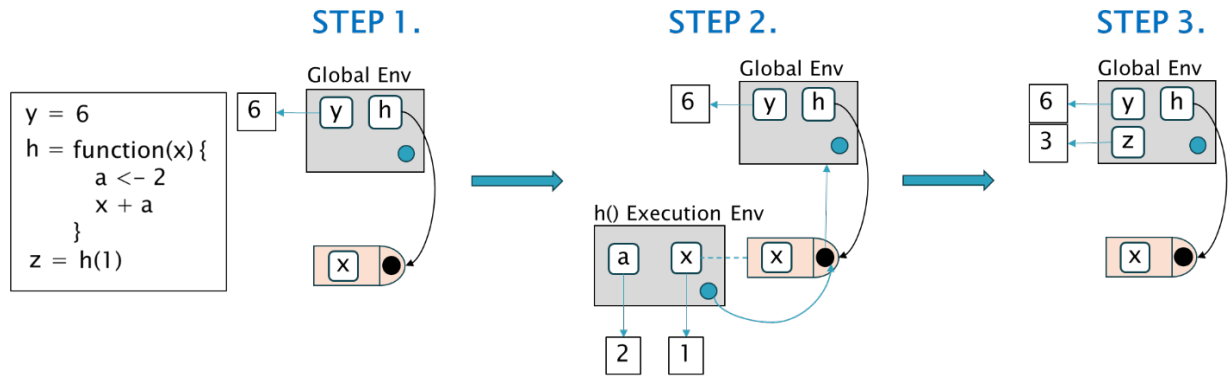

**Figure 3. Demonstration of environments in R**

Most environments have a higher-level environment, known as a parent, and even the global environment has a parent: the empty environment. Notably, the empty environment is the only one without a parent (Wickham, 2019). Understanding the parent of an environment is crucial because it determines where the environment will look for the value of a free variable. Free variables are those used within a function that are neither formal arguments nor local variables defined inside the function's body. The parent of a function's execution environment is determined by the location where the function is defined. For example, in Figure 3, the parent of h() execution environment is the global environment, because function h() is defined in global environment.

## 4. Scoping rules in SAS and R

### 4.1 Dynamic scoping in SAS

In SAS, the scope of macro variables is determined by whether they are defined in a global or local context. By default, any macro variable created outside of a macro definition automatically resides in the Global symbol table. Within a macro definition, the scoping rules prioritize existing local variables first, which means the newly defined macro variables are local variables inside a macro. If a macro has parameters, those variables are automatically local to the macro. SAS will search for a macro variable within the local symbol table first, and if not found, it will look "upwards" through any calling macros, ultimately checking the Global symbol table as a last resort.

Now, you probably have already known how the scoping works in SAS, would you please help me to predict the value of x1 and y1 after running all codes in SAS program 5?

```
/*example of dynamic scoping*/
%let x1=9;
%let y1=7;
%macro my_macro2;
%let x1=19;
%let y1=17;
%mend;

%my_macro2()
```

**SAS program 1 Example 1 of dynamic scoping**

The answer for this test is: x1=19 and y1=17, some people may find out the answer is surprising, why is SAS using the "local" variables assigned inside a macro to replace the global ones? Or our question can be: while SAS macro is executing, are all macro variables defined inside the macro local ones? The answer is obvious: no, not all the time. In this case, while we are assigning 19 to x1 in my_macro2, SAS will search in the local symbol table for x1 first. If x1 is not found in the local symbol table, SAS then will search in the global symbol table. Since x1 has already been in the global symbol table, in this case, SAS will directly replace the value of x1 with 19 in the global symbol table.

Let's delve deeper to understand what's happening behind the scenes. Unlike most modern programming languages that use **lexical scoping** (also known as static scoping), SAS generally utilizes **dynamic scoping** as its scoping rule (Ling & Wang, 2025). On the one hand, lexical scoping will look up values of variables based on the lexical

(textual) content defined by where the variable or function is called. On the other hand, in dynamic scoping, if a variable's scope is a macro, then its scope is the time-period during which the macro is executing. This means the determination of the variable's assignment relies on the runtime context, making it challenging to predict. Since dynamic scoping happens during the execution after the compilation, neither the macro processor nor compiler can tell whether an assignment statement is meant to assign a local macro variable or a higher scope macro variable that may exist at runtime (Joe, 2016).

Let's see a more straightforward example for dynamic scoping, illustrated in Figure 4. On the left, we have example code demonstrating dynamic scoping with a nested macro structure. Two macros, %inner and %outer, are defined globally, with %inner being called inside the %outer. To clarify the process, a graph is used to illustrate the mechanics behind the scenes. There are two instances of x defined : one in the global scope and another in the local scope of %outer, which are stored in the global symbol table and local symbol table for %outer, respectively. When SAS resolves the value for &x in %inner, it finds there is no local definition of x within %inner(). Thus, it searches for x in the scope of its runtime. Because the %inner is called inside of %outer, which means the macro inner is running in the outer, SAS will use x=6 defined in outer as the value to put in the log.

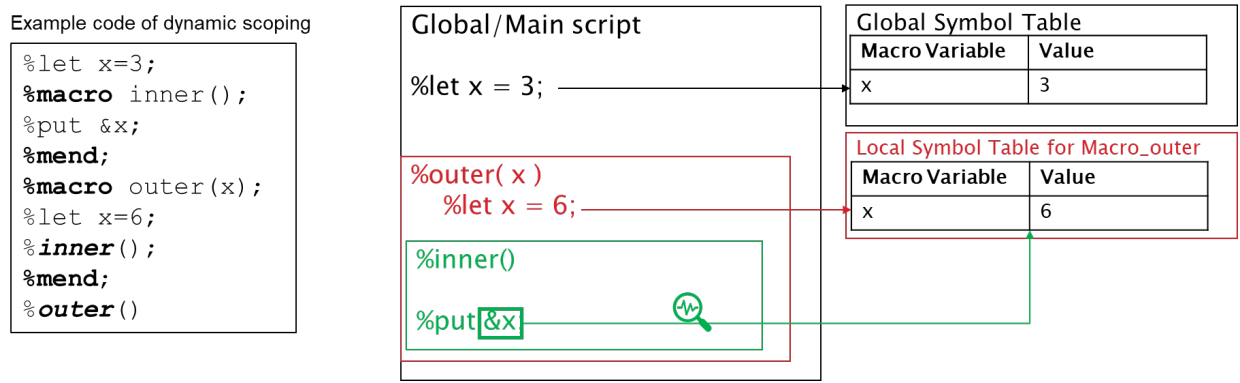

**Figure 4. Example of dynamic scoping in SAS**

## 4.2 Lexical Scoping in R

To understand the scoping rules of R, we have to know how R is binding values to the free variables. If the value of a variable cannot be found inside the function, then it will look at the environment in which the function is defined (Peng, 2022). If the search fails in this environment, it will continue to search in the parent environment. The process continues through successive parent environments until reaching the top-level environment, which is typically either the global environment (workspace) or a package's namespace. Beyond the top-level environment, the search persists through the search list until it encounters the empty environment.

Next, let's see an example for better understanding, as shown in Figure 5. Within the execution environment of function f3(), the variable y is set to x*2, however, x is not in its environment. To find the value of x, the search begins in f3()'s parent, f2(). Since x is not found in f2()'s environment, the search will continue up the "parent chain" to the top level. Luckily, we found an x in global environment, so y will be assigned with 4 (2*2).

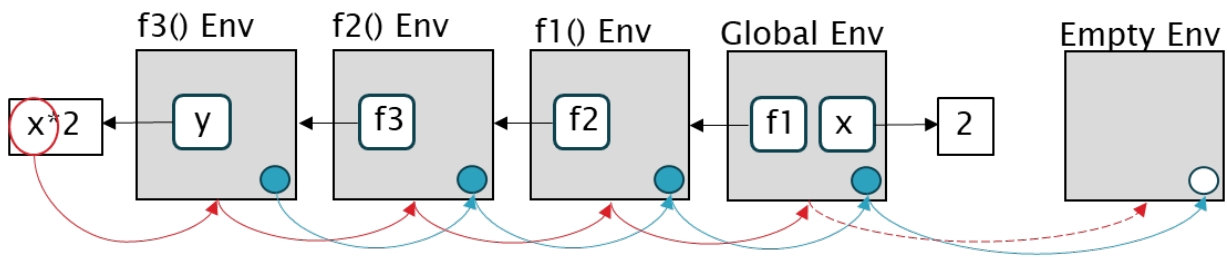

**Figure 5. Demonstration of lexical scoping in R**

Just like most modern programming languages, R utilizes lexical scoping. Lexical scoping does not consider when the function is called, it's only based on where the function is defined. Let's revisit the example from Figure 4; below is Figure 6. On the left, you'll find example code with a similar structure but yielding different results. On the right, a graphical demonstration of the code is presented. Since outer() and inner() are both defined in global environment, the global environment is the parent for both of them. When inner() is called inside outer(), it first searches for x in its local environment. Since x is not defined there, it proceeds to search in the parent of inner(), the global environment, rather than the environment of outer(), as would happen in SAS. Finally, it will find the x defined as 3 in the global environment and print it out.

Example code of lexical scoping

```
x=3
inner<-function(){
   print(x)
}
outer<-function(){
   x=6
   inner()
}
outer()
```

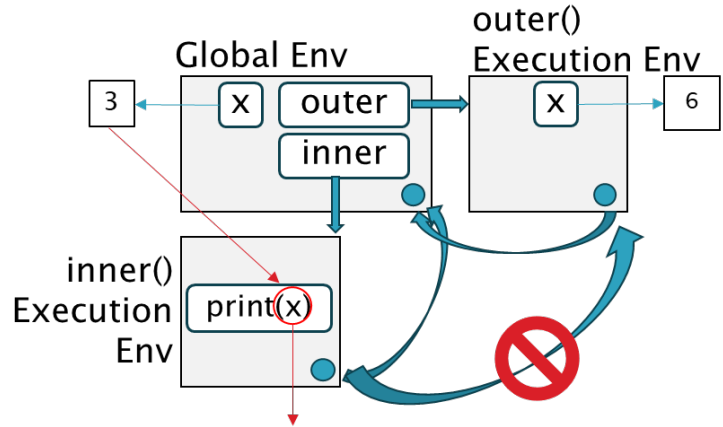

**Figure 6. Example of lexical scoping in R**

## 5. Inspect variables in a scope

### 5.1 Check symbol tables in SAS

Debugging and verifying the placement of macro variables is crucial for ensuring the accuracy of SAS programs. The %PUT statement with the _user_ keyword can be used to write a list of all user-defined macro variables, their values, and their respective scopes to the SAS log. In SAS program 2, %put _user_ is used to print the global and local symbol tables in the log. As shown in Figure 1, there are some extra variables in the global symbol table, please note that those are SAS system generated variables, we can ignore them most of the time.

```
/*example of checking symbol tables*/
%let y=8;
%let z=9;
%macro lazy(x=1,y=%eval(&x*10),z=%eval(&a+&b));
%let x=2;
%let a=3;
%let b=4;
%put (&x &y &z);
%put _user_;
proc print data=sashelp.vmacro;
run;
%mend;

%lazy()
```

**SAS program 2 Example of checking symbol tables**

```
(2 20 7)
LAZY A 3
LAZY B 4
LAZY X 2
LAZY Y %eval(&x*10)
LAZY Z %eval(&a+&b)
GLOBAL SASWORKLOCATION "/op
GLOBAL SYSSTREAMINGLOG true
GLOBAL Y 8
GLOBAL Z 9
```

**Figure 7 Local and global symbol tables printed in log**

Alternatively, the sashelp.vmacro view can be accessed, which provides a comprehensive dataset of all macro variables along with their names, values, and scopes, allowing users to directly inspect and manage both local and global symbol tables effectively.

| Obs | scope | name | offset | value |
|---|---|---|---|---|
| 1 | LAZY | A | 0 | 3 |
| 2 | LAZY | B | 0 | 4 |
| 3 | LAZY | X | 0 | 2 |
| 4 | LAZY | Y | 0 | %eval(&x*10) |
| 5 | LAZY | Z | 0 | %eval(&a+&b) |
| 6 | GLOBAL | SASWORKLOCATION | 0 | "/opt/SASWORK/F |
| 7 | GLOBAL | SYSSTREAMINGLOG | 0 | true |
| 8 | GLOBAL | Y | 0 | 8 |
| 9 | GLOBAL | Z | 0 | 9 |

Figure 8 The content of sashelp.vmacro

## 5.2 Check objects in enviroments

In R, we can use ls() or objects() to return a vector of character strings giving the names of the objects in the environment. To obtain both the names and their associated values, you can use the ls.str() function, which can return the object class along with its value for variables or structure for datasets.

Another way of checking the objects in global environment is to check the environment tab (Figure 9) in RStudio, in this tab you will find all objects in the global environment.

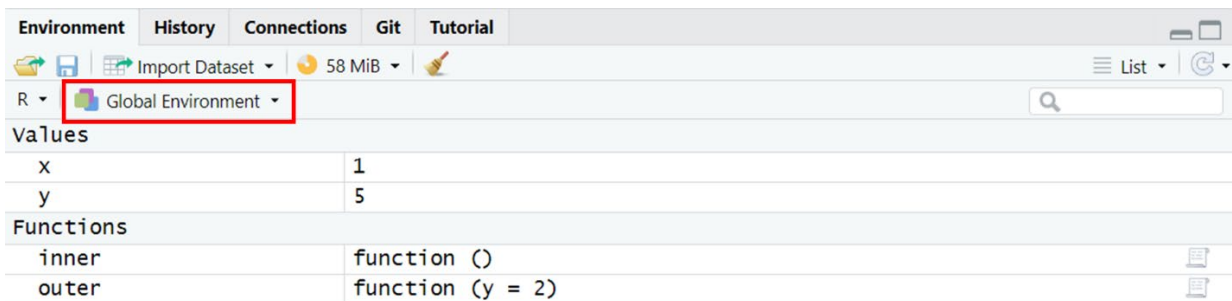

Figure 9. Environment tab in RStudio

## 6. Controlling variable scope

### 6.1 Controlling variable scope in SAS

As I mentioned, SAS macro "generally" employs dynamic scoping, however, there are ways to explicitly specify the macro variables to be local which can enforce lexical scoping. In SAS program 3, x2 is used as the parameter in the macro, which can automatically make it a local variable. SAS also provides the %local statement, which can turn y2 into a local macro variable, and since x2 and y2 are both local variables, the value of global x2 and y2 will not change.

%local statement ensures that a macro variable is confined to the local environment of the macro where it is declared. If the macro variable already exists, %local will be ignored; otherwise, it will create a macro variable in the local symbol table. The good practice is that we should always use %local to specify variables defined inside a macro to avoid any conflict.

In addition, we can use %global to explicitly specify a variable inside a macro to be global variable. However, for a multi-layer nested scoping structure, you cannot force a variable to be in an intermediate scope between the global scope and most local scope.

```
/*example of explicitly specified local macro variables*/
%let x2=9;
%let y2=7;
%macro my_macro2(x2);
%local y2;
```

```
    %let x2=19;
    %let y2=17;
    %mend;

%my_macro2()
```

**SAS program 3 Example of explicitly specified local macro variables**

### 6.2 Controlling variable scope in R

In R, the <- operator is used to assign a value to a variable in the current environment. In situations where you might need to assign value in a parent environment, the super assignment operator <<- is used. Alternatively, you can control the scope by assigning a value to a variable in the global environment using .GlobalEnv$variable <- value, similar to %global in SAS.

However, these methods are generally discouraged because they can produce side effects by altering variables outside their intended scope, beyond what the function explicitly returns. This can lead to unexpected issues, especially for users unfamiliar with the functions involved. For instance, an unexpected outcome might overwrite existing variables in the calling environment, potentially impacting subsequent code. It's recommended to avoid using these scope-controlling techniques in programming to maintain clarity and prevent unintended consequences.

## 7. Conclusion

In conclusion, variable scoping is a critical and foundational element of functional programming, significantly impacting how variables are accessed and managed in SAS and R. This paper compared the dynamic scoping of SAS, which offers runtime flexibility through symbol tables, with the lexical scoping of R, providing stability and predictability via environments. Through illustrative code examples, the paper highlighted how these differences in scoping rules can lead to varied outcomes in code execution, deepening understanding of their practical implications. The practical methods for inspecting and controlling variable scopes further empower programmers to enhance code precision and reliability. This understanding is especially valuable for developers transitioning between these languages, as it equips them with the knowledge to optimize their coding practices effectively. Ultimately, a deep comprehension of these scoping rules aids in achieving more accurate programming outcomes and a smoother transition in environments where both SAS and R are used.